# Synthesis and Characterization of Superparamagnetic Iron Oxide Nanoparticles: A Series of Laboratory Experiments

Armando D. Urbina[1+], Hari Sridhara[1+], Alexis Scholtz[2+], Andrea M. Armani[1,2,3] *

[1] Mork Family Department of Chemical Engineering and Materials Science, University of Southern California, Los Angeles, CA 90089, USA

[2] Alfred E. Mann Department of Biomedical Engineering, University of Southern California, Los Angeles, CA 90089, USA

[3] Ellison Institute of Technology, Los Angeles, CA 90064, USA

[+] These authors contributed equally.

*aarmani@eit.org

## ABSTRACT

The following laboratory procedure provides students with a hands-on experience in nanomaterials chemistry and characterization. This three-day protocol is easy to follow for undergraduates with basic chemistry or materials science backgrounds and is suitable for inclusion in upper division courses in inorganic chemistry or materials science. Students use air-free chemistry procedures to synthesize and separate iron oxide magnetic nanoparticles and subsequently modify the nanoparticle surface using a chemical stripping agent. The morphology and chemical composition of the nanoparticles are characterized using electron microscopy and dynamic light scattering measurements. Additionally, magnetic characterization of the particles is performed using an open-source (3D-printed), inexpensive magnetophotometer. Possible modifications to the synthesis procedure including the incorporation of dopants to modify the magnetic response and alternative characterization techniques are discussed. The three-day synthesis, purification, and characterization laboratory will prepare students with crucial skills for advanced technology industries such as semiconductor manufacturing, nanomedicine, and green chemistry.

## GRAPHICAL ABSTRACT

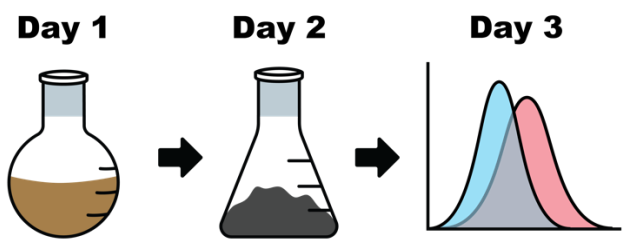

## INTRODUCTION AND BACKGROUND

In 2000, the National Nanotechnology Initiative was launched in the United States with four objectives designed to advance workforce training and technology development.[1] Since then, the research community has made significant discoveries in nanotechnology research (first objective). However, the development and sustainment of educational resources to support the creation of a skilled workforce (third objective) has not received similar attention.[2] As discussed in several longitudinal studies dating back to the mid-2000's, many undergraduate and graduate chemistry and chemical engineering laboratory experiments have not been updated to include recent technical advances and modern instrumentation.[3–5] In addition to decreasing the utility of the degrees, this also has resulted in increased attrition from Science, Technology, Engineering, and Mathematics (STEM) degree programs.[6] In this context, the development of nanomaterial synthesis laboratories provides a clear strategy to increase engagement and retention in a wide range of degree programs.[7]

There are several examples of inquiry-based undergraduate teaching laboratories exploring fundamental nanoengineering concepts[8–11] and training students in various nano-characterization techniques.[12,13] One family of nanomaterials commonly incorporated into teaching laboratories are optically-responsive nanoparticles, including gold nanoparticles[14–18] and fluorescent quantum dots.[11,19,20] While these materials serve as an ideal platform for exploring complex quantum concepts, such as the discretization of energy states[21,22] and manipulation of light at the nanoscale,[23,24] the notable properties of these nanoparticles are limited to geometry and optical response. In contrast, nanomaterials can demonstrate a plethora of behaviors, including mechanical,[25] optical,[2,23,24] electrical,[26] chemical,[27] and more.[28–31] One unique nanomaterial is iron oxide nanoparticles. Unlike quantum dots or metal nanoparticles, iron oxide nanoparticles exhibit size- and shape-tunable magnetic behavior.[27,32–34] As a result of their magnetic response and low biotoxicity, this nanomaterial shows promise in a wide range of applications and fields, including biomedical devices and imaging, energy storage and generation, and chemical processing.[28,31,35–37] Thus, given their synergistic relationship to existing undergraduate nanoparticle laboratories and their emerging real-world applications, the synthesis and characterization of iron oxide nanoparticles provide a rich foundation for integration into an academic laboratory course.[38]

Iron oxide nanoparticles fall under a larger class of materials known as paramagnetic materials. Unlike many commonly available magnetic materials which rely on collective, long-range magnetic order to achieve an intrinsic magnetic response, paramagnetic and diamagnetic materials are considered nonmagnetic in the absence of an external magnetic field.[39] However, in the presence of a field, their magnetic domains align in a single direction (Figure 1), and the particle exhibits a magnetic response. The electron pairing determines if a material is diamagnetic or paramagnetic, which governs the sign of the magnetic susceptibility. Symmetric or spherical iron oxide nanoparticles demonstrate a paramagnetic response due to their unpaired electrons and have a positive magnetic susceptibility.[33] This response is governed by the material composition, crystallographic uniformity, and size. Indirectly, it is also controlled by the sample preparation, as clusters of particles behave differently from well-dispersed particles. Given the diversity of applications for paramagnetic iron oxide nanoparticles, it is useful to measure their size, composition, and magnetic response before pursuing an application.[40–42]

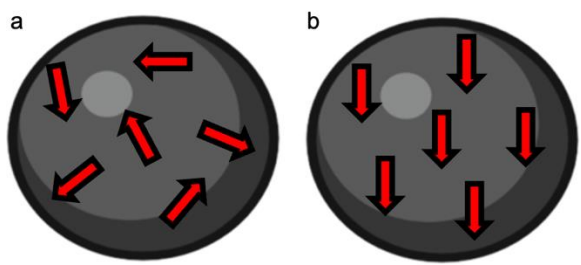

Figure 1. A paramagnetic nanoparticle (a) before and (b) during exposure to an external magnetic field. This figure was prepared using Biorender.

In prior work, several methods have been used to characterize these properties in iron oxide nanoparticles specifically. The methods and their generated data are presented in Table 1. As can be observed, there are several options for measuring each parameter, depending on the availability of instrumentation. Additionally, depending on the method(s) selected, the results will be provided at either the single particle level or at the population level. This provides a unique opportunity to compare the difference between single particle and population-level data.

**Table 1. Parameters of Interest and Characterization Methods**

| Parameter Measured | Measurement Method | Data Visualization | Type of Data |
|---|---|---|---|
| Particle size | Dynamic Light Scattering (DLS)[43] | Histogram | Population level |
| Particle size | Transmission Electron Microscope (TEM)[44] | Image | Particle level |
| Particle size | Scanning Electron Microscope (SEM)[45] | Image | Particle level |
| Particle size | Atomic Force Microscopy (AFM)[46] | 2.5D image | Particle level |
| Dopant concentration | Energy Dispersive X-ray Spectroscopy (EDX/EDS)[47] | Spectra | Population level |
| Dopant concentration | X-Ray Diffractometry (XRD)[48] | Spectra | Population level |
| Magnetic response | Magnetophotometry (MAP)[49] | Graph | Population level |
| Magnetic response | SQUID[50] | Graph | Population level |
| Magnetic response | Magnetic Resonance Imaging (MRI)[51] | Image | Population level |

## EXPERIMENTAL DETAILS

The laboratory protocol is broken into 3 days: nanoparticle synthesis on Day 1, sample preparation and nanoparticle surface treatment on Day 2, and nanoparticle characterization on Day 3. A general overview is illustrated in Figure 2, and the detailed protocol is contained in the supplementary Instructor and Student Manuals.

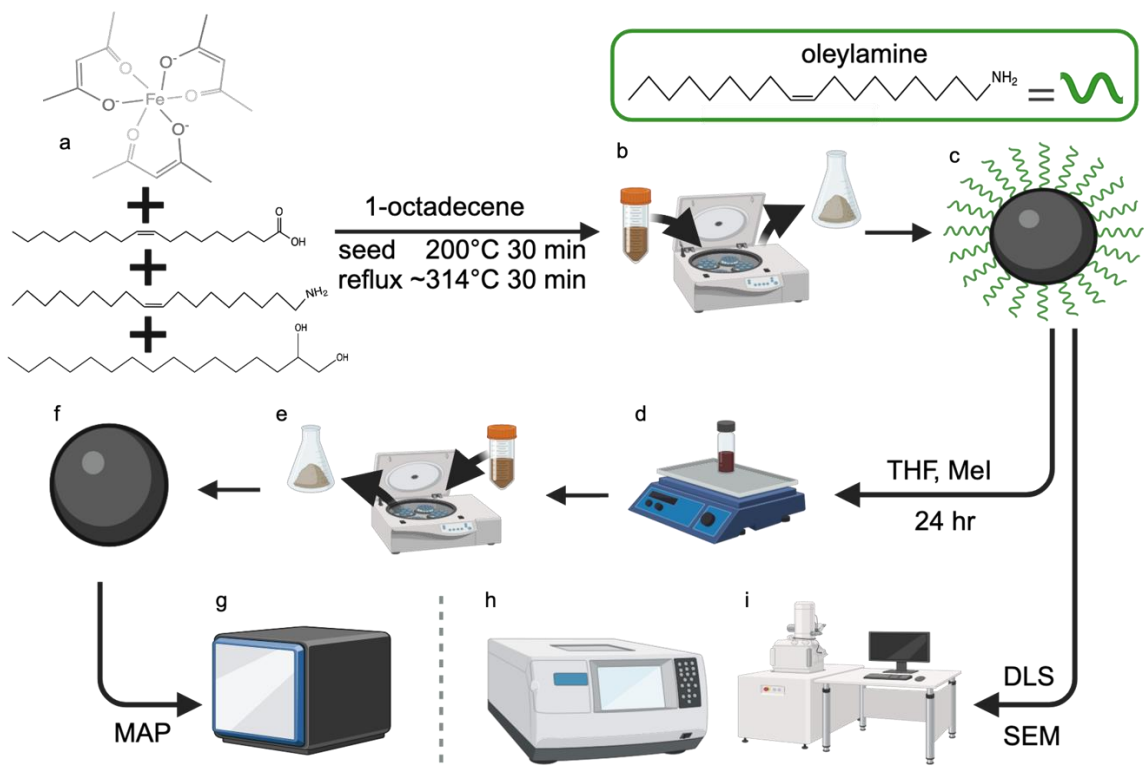

Figure 2. Overview of the full laboratory procedure. (a-c) Day 1 consists of nanoparticle synthesis including (a) the $Fe_3O_4$ synthesis reaction, (b) isolating the iron oxide particles by centrifuge, and (c) drying the iron oxide nanoparticles coated with oleylamine ligands. (d-f) Day 2 includes (d) the stripping of nanoparticles by iodomethane (MeI), (e) centrifugation to isolate the stripped iron oxide nanoparticles and (f) drying the stripped $Fe_3O_4$ nanoparticles. (g-i) Characterization techniques used in Day 3 include (g) magnetophotometry (MAP) using stripped nanoparticles, (h) Scanning Electron Microscopy (SEM) using oleylamine coated nanoparticles, and (i) Dynamic Light Scattering (DLS) using oleylamine coated nanoparticles. This figure was prepared using BioRender.

Single-crystal metal nanoparticles formed via a two-part synthesis of nucleation and growth steps are well understood to produce monodisperse size distributions with strong shape control.[52–56] The one-pot two-part synthesis used here follows a previously published procedure with slight modifications to align with undergraduate laboratory expectations.[57,58] Briefly, solid and liquid reagents are combined with the reaction solvent in an air-free environment (Figure 2a).[59] Holding the reagents at an initial temperature of 200°C cleaves the acetylacetone bonds in Fe(acac)$_3$, allowing for nucleation of iron oxide sites. Further increasing the temperature until reflux, when particle growth is thermodynamically preferential, for a predetermined time leads to single crystal and highly monodisperse particles (Figure 2b).

After purification, a subset of particles is further treated with iodomethane (MeI) to remove or 'strip' the oleylamine ligands attached to the particle surface (Figure 2c). Samples are then prepared with the untreated and stripped particles (Figure 2d-f) to conduct size, shape, and magnetic response characterization (Figure 2g-i). This student cohort used dynamic light scattering (DLS), scanning electron microscopy (SEM), and an open-source magnetophotometer (MAP)[49] to analyze the particle size, shape, and magnetic response. Possible variations to the synthesis and alternative characterization methods are listed in Table 1 and are discussed in more depth in the Instructor Manual.

## Hazards and Safety

A concrete understanding of laboratory safety is essential for pursuit of modern careers in science and engineering. This laboratory procedure involves use of numerous hazardous chemicals and requires the use of appropriate mitigation measures. Before the start of the laboratory, all students received general laboratory safety training, which included instruction in proper handling of chemicals, personal protective equipment (PPE), mitigation measures (location and use of eye-wash stations, chemical showers, and fire extinguishers), and identification of chemicals by hazard type. During the laboratory, proper PPE (high-temperature lab coat, goggles, and nitrile or heat-resistant gloves) was used in accordance with standard chemistry practices. All chemistry procedures were conducted in a fume hood to limit airborne exposure, and a trained laboratory assistant was present throughout the procedure to assist when necessary. Additional hazard information is included in the Instructor and Manuals.

## RESULTS AND DISCUSSION

### Educational cohort

A cohort of students from the University of Southern California Viterbi School of Engineering with a minimum requirement of at least one semester equivalent of general chemistry volunteered to complete the lab to increase the diversity of their undergraduate research experiences. The students ranged from first-year through fourth-year undergraduates pursuing a range of degree programs. All students had completed general laboratory safety training.

Under the supervision of the laboratory assistant, the students worked in groups of 2-3 over the course of the three-day lab, and each day was typically 3-4 hours. The experiments were designed to support inquiry-based learning by encouraging the students to discover the answers to a series of scientific questions each day as well as expand their technical capabilities. These learning objectives can be summarized as follows:

1. Calculating the amount of reagents required using stoichiometric principles.
2. Practicing commonly used chemistry skills, including centrifugation.
3. Introducing students to laboratory techniques related to air-sensitive chemistry, such as using a glove box and Schlenk line.
4. Executing various sample preparation protocols.
5. Learning and practicing nanomaterial characterization techniques.
6. Applying theoretical knowledge of characterization techniques and assessing the characteristics of the synthesized nanoparticles.
7. Developing data visualization and scientific communication skills.

Because this laboratory introduced new techniques and skills, the students were instructed in the principles behind each piece of equipment and executed the protocol themselves but were still closely monitored by a laboratory assistant at each step of the procedure (objectives 1-3). After carrying out the synthesis, students prepared samples for characterization according to the instructions listed in the Student Manual (objectives 4-5). Students performed all data analysis for subsequent discussion (objectives 6-7).

To assess if these learning objectives were achieved, the students' understanding of the experimental and scientific principles was evaluated throughout the three days using diagnostic assessment, including pre-tests, and formative assessment, including passive and active observational techniques. For example, throughout the experiments, the laboratory assistant asked questions to assess understanding. The pre-tests and solutions as well as example questions are included in the supplementary documentation.

Immediately following the conclusion of the laboratory, a summative assessment was performed, which included both qualitative and quantitative components. Notably, 100% of the students felt that the lab experiment should be integrated as a required element into the undergraduate curriculum, and 100% felt that they gained valuable skills and/or knowledge applicable outside of this lab session. This strong, positive response is notable given the students' diverse academic backgrounds. However, the cohort had a mixed response when asked about the level of the rigor of the lab. Given the wide range in course preparation and academic levels, this response is not surprising. To address this possible shortcoming, several variations are included in the Instructor Manual to increase the complexity.

In addition, a secondary reflective assessment was performed six months later. Upon reflection, the primary skills and knowledge imparted are the new data analysis methods and visualization techniques and the instrumentation methods. The students indicated the value of these as they relate to their ongoing coursework and their intended future career paths. It is notable that this group of students have already found value in this new knowledge and anticipate or are already applying it across a diverse range of fields, spanning from machine learning to therapeutics and biomedical engineering. This rapid impact highlights the relevance of this lab to an undergraduate engineering degree program. The entire assessment is included in the Instructor Manual.

## Nanoparticle Characterization

To confirm the suitability of the synthetic protocols and the proposed characterization methods for an undergraduate laboratory setting, Figure 3 presents a series of results from nanoparticles prepared by the laboratory assistant and from particles prepared by a single student group. Both sets of particles are synthesized following the procedure included in the Student Manual. Due to limitations in equipment access, the laboratory assistant performed the DLS measurements, but the students took SEM images and performed magnetic response testing with a MAP themselves. The students performed all sample preparation and data analysis. For all data sets in Figure 3, the results from the student group agreed with those from the laboratory assistant, demonstrating the robustness of the synthetic method.

The combination of SEM and DLS data provides an opportunity to analyze particle size at both the single particle and ensemble level. Based on published work by Sun et al., we anticipate the particle size to be between 10-15 nm.[57] Each group prepared samples and acquired SEM images of their own batch of nanoparticles as seen in Figure 3a. Notably, the image shows particle clumps in the 100 nm range and individual particles are not clearly resolved (Figure 3a). For the particles synthesized by the laboratory assistant, the diameter determined from SEM images is in the 15 nm range (Figure 3b). The DLS data reveals similar particle sizes, diameters of 12 nm, for both the student and the laboratory assistant samples (Figure 3c-d). Lastly, students confirmed the paramagnetic response of the particles using a custom Magnetophotometer (MAP) system (Figure 3e-f). As can be observed in Figure 3e and 3f, the particles synthesized by either the students or the laboratory assistant exhibit a clear positive magnetic response when in the presence of the permanent magnetic, indicating a positive magnetic susceptibility value.

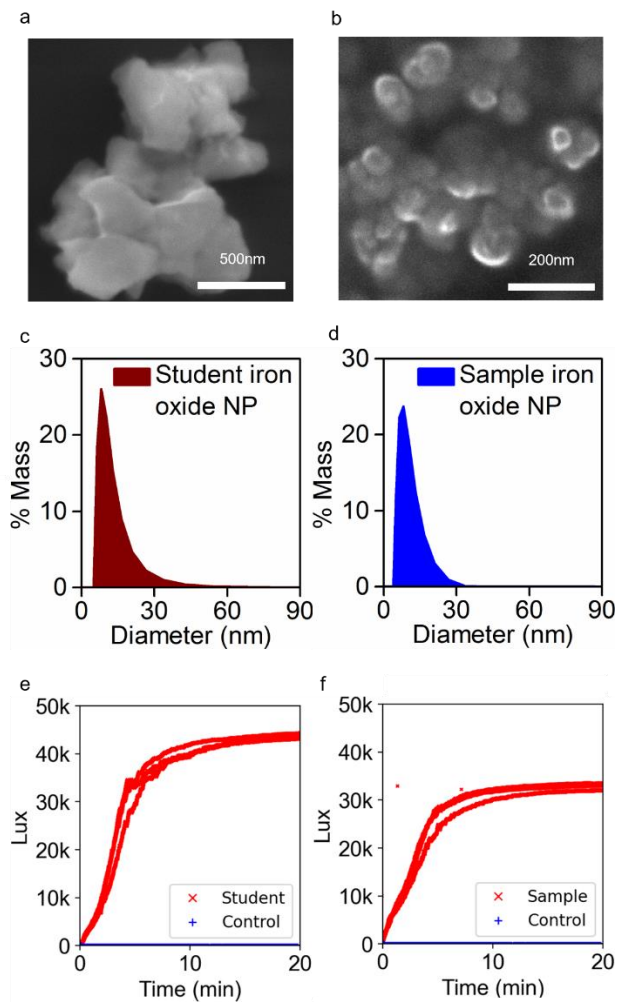

Figure 3. Iron oxide nanoparticle (a-b) SEM image, (c-d) DLS data, (e-f) magnetic response data from the MAP system and ethanol control. Data taken with particles made by the student volunteers following the procedure are shown in panels (a), (c), and (e). Data taken with particles made by graduate assistant are shown in panes (b), (d), and (f).

## SUMMARY AND CONCLUSION

In summary, we have detailed a research-based, three-day, advanced undergraduate laboratory protocol which includes the synthesis of iron oxide magnetic nanoparticles, modification of the surface coating, and characterization of the particle properties. The experiments can easily be modified based on available equipment. They are designed to combine principles from chemistry, materials science, and physics to expose students to interdisciplinary research and modern instrumentation while utilizing inexpensive reagents and commonly accessible characterization techniques. They also serve as a bridge, connecting seemingly disparate coursework. While this 3-day laboratory experiment was piloted with a cohort of undergraduate students, it could also be suitable for a graduate student experimental methods course.

## ASSOCIATED CONTENT

### Supporting Information

The Supporting Information is available on the ACS Publications website at DOI: 10.1021/acs.jchemed.XXXXXXX.

Instructor Manual (PDF, DOCX): Detailed protocols, trouble-shooting suggestions, materials and equipment tables, and educational assessment.

Student Manual (PDF, DOCX): Step-by-step protocols and materials and equipment tables

Student Assessment (PDF, DOCX) and Solutions (PDF, DOCX): Suggested pre-laboratory exercises to enrich student learning and solutions.

## AUTHOR INFORMATION

Corresponding Author: Andrea M. Armani

Email: aarmani@eit.org

Armando Urbina: 0000-0003-3036-4991

Hari Sridhara: 0000-0001-8548-0581

Alexis Scholtz: 0000-0001-6275-0622

Andrea M. Armani: 0000-0001-9890-5104

## ACKNOWLEDGMENTS

The authors would like to thank Jack G. Paulson, Kylie J. Trettner, Marko Lilić, Patrick Saris, Sydney Fiorentino, Victoria Nuñez (University of Southern California) for contributions to the refinement of the iron oxide nanoparticle protocol. This work was supported by the Office of Naval Research (N00014-22-1-2466, N00014-21-1-2044), the National Science Foundation (DBI-2222206), and the Ellison Institute of Technology, LLC. A. D. Urbina is supported by a National Science Foundation Graduate Research Fellowship. H. Sridhara is supported by the University of Southern California Center for Undergraduate Research in Viterbi Engineering and Provost Undergraduate Research Fellowships.

A.M.A. serves as the Senior Director of Engineering and Physical Sciences for the Ellison Institute of Technology, LLC (paid).